\documentclass[aps,prl,twocolumn,groupedaddress]{revtex4}
\usepackage{amssymb}
\usepackage{epsfig}
\usepackage{graphicx}

\begin{document}

\newcommand{\bra}[1]{\langle {#1} |}
\newcommand{\ket}[1]{| {#1} \rangle}
\newcommand{\expect}[1]{\langle {#1} \rangle}
\newcommand{\ketn}[1]{ {#1} \rangle}
\newcommand{\bea}{\begin{eqnarray}}
\newcommand{\eea}{\end{eqnarray}}
\newcommand{\beq}{\begin{equation}}
\newcommand{\eeq}{\end{equation}}
\newcommand{\lav}{\langle}
\newcommand{\rav}{\rangle}
\def \tr{{\mbox{tr~}}}
\def \ra{{\rightarrow}}
\def \ua{{\uparrow}}
\def \da{{\downarrow}}
\def \be{\begin{equation}}
\def \ee{\end{equation}}
\def \ba{\begin{array}}
\def \ea{\end{array}}
\def \bea{\begin{eqnarray}}
\def \eea{\end{eqnarray}}
\def \nn{\nonumber}
\def \l{\left}
\def \r{\right}
\def \half{{1\over 2}}
\def \etal{{\it {et al}}}
\def \cH{{\cal{H}}}
\def \cM{{\cal{M}}}
\def \cN{{\cal{N}}}
\def \cQ{{\cal Q}}
\def \cI{{\cal I}}
\def \cV{{\cal V}}
\def \cG{{\cal G}}
\def \cF{{\cal F}}
\def \cZ{{\cal Z}}
\def \bS{{\bf S}}
\def \bI{{\bf I}}
\def \bL{{\bf L}}
\def \bG{{\bf G}}
\def \bQ{{\bf Q}}
\def \bK{{\bf K}}
\def \bR{{\bf R}}
\def \br{{\bf r}}
\def \bu{{\bf u}}
\def \bq{{\bf q}}
\def \bk{{\bf k}}
\def \bv{{\bf v}}
\def \bx{{\bf x}}
\def \bpsi{{\bar{\psi}}}
\def \tJ{{\tilde{J}}}
\def \W{{\Omega}}
\def \e{{\epsilon}}
\def \lam{{\lambda}}
\def \L{{\mathcal L}}
\def \a{{\alpha}}
\def \t{{\tau}}
\def \b{{\beta}}
\def \g{{\gamma}}
\def \D{{\Delta}}
\def \de{{\delta}}
\def \w{{\omega}}
\def \r{{\rho}}
\def \s{{\sigma}}
\def \f{{\varphi}}
\def \x{{\chi}}
\def \e{{\epsilon}}
\def \h{{\eta}}
\def \G{{\Gamma}}
\def \z{{\zeta}}
\def \hatt{{\hat{\t}}}
\def \hn{{\bar{n}}}
\def \vk{{\bf{k}}}
\def \vq{{\bf{q}}}
\def \gk{{\g_{\vk}}}
\def \nd{{^{\vphantom{\dagger}}}}
\def \yd{^\dagger}
\def \d{{d\!\!\!^-}}
\def \av#1{{\langle#1\rangle}}
\def \ket#1{{\,|\,#1\,\rangle\,}}
\def \bra#1{{\,\langle\,#1\,|\,}}
\def \braket#1#2{{\,\langle\,#1\,|\,#2\,\rangle\,}}

\title{Nematic Order by Disorder in Spin-2 BECs}
\author{Ari M. Turner$^1$, Ryan Barnett$^2$, Eugene Demler$^1$, and Ashvin Vishwanath$^3$}
\affiliation{$^1$Department of Physics, Harvard University,
Cambridge, Massachusetts 02138, USA} \affiliation{$^2$Department of
Physics, California Institute of Technology, MC 114-36, Pasadena,
California 91125,USA} \affiliation{$^3$Department of Physics,
University of California, Berkeley, California 94720, USA}
\date{\today}

\begin{abstract}
The effect of quantum and thermal fluctuations on the phase diagram
of spin-2 BECs is examined. They are found to play an important role
in the nematic part of the phase diagram, where a mean-field
treatment of two-body interactions is unable to lift the accidental
degeneracy between nematic states. Quantum and thermal fluctuations
resolve this degeneracy, selecting the uniaxial nematic state, for
scattering lengths $a_4>a_2$, and the square biaxial nematic state
for $a_4<a_2$. Paradoxically, the fluctuation induced order is
stronger at higher temperatures, for a range of temperatures below
$T_c$. For the experimentally relevant cases of spin-2 $^{87}$Rb and
$^{23}$Na, we argue that such fluctuations could successfully
compete against other effects like the quadratic Zeeman field, and
stabilize the uniaxial phase for experimentally realistic
conditions. A continuous transition of the Ising type from uniaxial
to square biaxial order is predicted on raising the magnetic field.
These systems present a promising experimental opportunity to
realize the `order by disorder' phenomenon.

\end{abstract}

\maketitle
\date{\today}

 The ground-state properties as well as the dynamics of spinor
Bose-Einstein condensates has been the subject of many experimental
studies over the past decade (see, for instance,
\cite{stenger98,schmaljohann04,chang04,griesmaier05,sadler06}). The
spin degree of freedom introduces a rich variety of spin ordered
superfluids \cite{ho98,ohmi98,ciobanu00} and the interaction between
spin order and superfluidity has several interesting consequences,
for example topological defects of the magnetic texture that trap
vorticity \cite{zhou01}. The precise nature of the spin order
realized in the ground state depends on the spin dependent two body
interaction between atoms, parameterized by the scattering length in
different total-spin channels. Usually, a specification of these
scattering lengths uniquely fixes the spin configuration of the
condensate. An interesting exception occurs in the case of spin two
atoms. Here, depending on the relative values of the scattering
lengths in the total spin 0, 2 and 4 channels ($a_0$,$a_2$,$a_4$),
one obtains either a ferromagnetic state, with a net spin moment, a
cyclic (or ``tetrahedratic'') state, which breaks time reversal
symmetry but does not have a net moment, or a nematic state which
preserves time reversal invariance. In contrast to the first two
states, the nematic state requires specifying an additional
parameter for its description. To visualize the parameter
geometrically, the state may be represented by its ``reciprocal
spinor,'' a configuration of four points. The nematic states have
the symmetry of a rectangle, and the additional parameter $\eta$ is
related to the aspect ratio of this rectangle. This parameter is not
fixed by the two body interactions, and a manifold of accidentally
degenerate states remains at this level \cite{barnett06}.

\begin{figure}
\includegraphics[width=.47\textwidth]{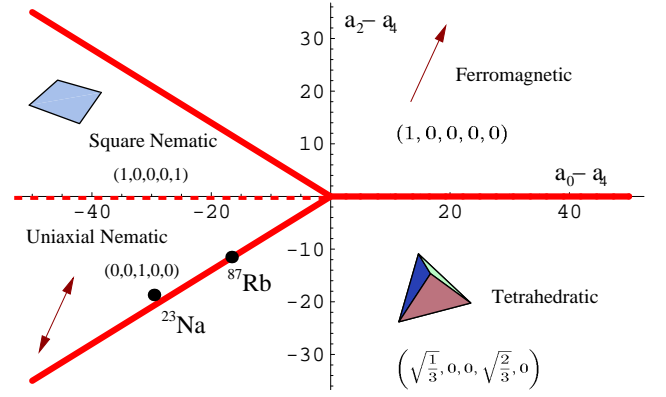}
\caption{The phase diagram for a spin-two spinor
condensate. The dashed line corresponds to the additional phase
boundary dividing the nematic phase into two phases, a uniaxial
nematic phase and a square nematic phase.}
\label{fig:peace}
\end{figure}

We show here that this accidental degeneracy is
lifted by superfluid phonons, whose velocities depend on the precise
nematic state being realized. In the low temperature regime, the
quantum zero point energies of the superfluid phonons leads to an
energy splitting between different nematic states. This is
reminiscent of the Casimir force arising from photon zero point
energies \cite{casimir}; here the `force' leads to a splitting of
nematic states. At higher temperatures, thermal fluctuations lift the
degeneracy in the free energy via entropic effects. Remarkably, we
are able to derive closed form analytic expressions for this
splitting in both these limits, yielding the phase diagram shown in
Fig.~\ref{fig:peace}. Fluctuations select the two high symmetry
states, the uniaxial nematic (with the symmetry of a line) and the
square biaxial nematic (with the symmetry of the square) in the
parts of the phase diagram shown. The latter is an elusive phase
with a non-abelian homotopy group, that was long sought after in
liquid crystal systems, but appears naturally here. Two
experimentally relevant systems, spin-2 $^{87}$Rb and $^{23}$Na, are
also shown on this phase diagram. They are expected to realize the
uniaxial nematic phase. Bose condensates of spin-2 $^{87}$Rb have
been realized \cite{schmaljohann04} and for the parameters of
that experiment we find that the splitting between nematic states
from quantum fluctuations to be of order $0.3$ pK  while the
free energy splitting from thermal fluctuations can be as large as
$6$ pK. The latter should be readily observable in
experiments with careful control over the quadratic Zeeman splitting
which competes with this fluctuation mechanism and prefers a square
nematic state oriented in a plane orthogonal to the field.

An unusual feature of the fluctuation selection is that the order is
actually stronger at higher temperatures (sufficiently far below the
transition temperature), in contrast to one's intuitive expectation
of thermal effects favoring disorder. This leads to the prediction
that for sufficiently weak background fields, a transition is
expected on warming the system in the superfluid state, as
fluctuation induced ordering overpowers the external field. These
effects, both quantum and thermal, are expected to be stronger in
the $^{23}$Na. While such `order by disorder' mechanisms have been
widely discussed in the theory of frustrated magnetism
\cite{Villain,Henley}, we are not aware of an unambiguous
manifestation of this effect in experiments. The spin-2 nematic
condensates seem like a promising system to observe this intriguing
effect.

{\it Mean Field Analysis:} In the dilute limit, spin-two bosons will
interact with the pair-potential $ V({\bf r_1}-{\bf
r_2})=\delta({\bf r_1}-{\bf r_2}) (g_0 {\cal P}_0 + g_2 {\cal P}_2 +
g_{4}{\cal P}_4) $ where $g_F=4\pi \hbar^2 a_F/m$ and ${\cal P}_{F}$
projects into the total spin $F$ state.  Such a pair potential leads
to the interaction Hamiltonian which can be conveniently expressed
as:
\begin{equation}
{\cal H}_{int}=\int d^3 r \frac{\alpha}{2} (\psi^{\dagger}\psi)^2
+\frac{\beta}{2} \psi^{\dagger}\mathbf{F}\psi\cdot
\psi^{\dagger}\mathbf{F}\psi +\frac{\gamma}{2}(\psi^{\dagger}\psi_t)
(\psi_t^{\dagger}\psi) \label{eq:CYH}
\end{equation}
where $\psi$ is a five component vector operator whose component
$\psi_m({\bf r})$ destroys a boson in spin state $F_z=m$ at
position $\bf r$.
Here, $\psi_t$ is the time reversal of $\psi$, namely
$(\psi_{-2}^{\dagger},
-\psi_{-1}^{\dagger},\psi_{0}^{\dagger},-\psi_{1}^{\dagger},
\psi_{2}^{\dagger})$. These parameters are related to the original
ones by $ \alpha=\frac{3g_4+4g_2}{7},\nonumber \;
\beta=-\frac{g_2-g_4}{7},\nonumber \;
\gamma=\frac{1}{5}(g_0-g_4)-\frac{2}{7}(g_2-g_4). $ The total
Hamiltonian is then ${\cal H} = {\cal H}_0 + {\cal H}_{int}$; where
the free part is ${\cal H}_0 =  \int d^3r \frac{\hbar^2}{2m}|\nabla
\psi|^2 - \mu (\psi^{\dagger}\psi)$.

The mean-field phase diagram for such a model is shown in
Fig.~\ref{fig:peace} \cite{ciobanu00}.
The various extrema correspond to the
ferromagnetic, nematic, and tetrahedratic phases.  We will focus on
the region where the nematic phase is stabilized.
The general nematic state, up
to an overall phase is $\phi_\eta
=(\frac{\sin\eta}{\sqrt{2}},0,\cos\eta,0,\frac{\sin\eta}{\sqrt{2}})$
and requires specification of the additional parameter $\eta$. As
$\eta$ is varied, the rectangle moves through each of the three
planes $xy$, $yz$, $xz$. $\eta=\frac{n\pi}{3}$ corresponds to a
uniaxial along an axis, while $\eta=(\frac{n}{3}-\frac{1}{2})\pi$
corresponds to a square in one of the coordinate planes.
At this level, the
different nematic states are degenerate but as demonstrated below,
fluctuations will remove this accidental degeneracy in the nematic
phase, producing the dashed phase boundary at $a_2=a_4$ ($\beta=0$).
Along this phase boundary however, the Hamiltonian possesses SO(5)
symmetry and hence the various nematic phases are exactly
degenerate. The mean-field energy density in the nematic phase is $
E_0/V = -\frac{\mu^2}{2(\alpha+\gamma)}$, while the chemical
potential itself is given by $\mu = (\alpha+\gamma)n_0$, where $n_0$
is the atom density. We now analyze the harmonic fluctuations about
the different nematic ground states.

{\it Fluctuations:} We follow the standard Bogoliubov theory of the
weakly interacting Bose gas to derive the spectrum of fluctuations.
Consider decomposing the field operator into a dominant c-number
piece, the mean field expectation value, and a smaller fluctuation
piece $\psi(r) = \sqrt{n_0}\phi_\eta + \delta\Psi(r)$. Substituting
this in the Hamiltonian we expand to quadratic order in the
fluctuations (linear terms vanish on choosing the mean field
solution) and diagonalize the resulting Bogoliubov Hamiltonian
${\cal H}_B$. This is aided by defining new canonical bosons via the
linear combinations: $ b_1\! =\! -\frac{i}{\sqrt{2}}(\delta\Psi_1 +
\delta\Psi_{-1})$, $b_2 \!=\! -\frac{1}{\sqrt{2}}(\delta\Psi_1 -
\delta\Psi_{-1})$, $b_3\! =\! \frac{i}{\sqrt{2}}(\delta\Psi_2 -
\delta\Psi_{-2})$, and $f_{\pm}\! =\! \frac1{\sqrt{2}}\delta\Psi_0
\pm \frac{i}{2}(\delta\Psi_2+\delta\Psi_{-2}) $. Further, we define
$p = \frac1{\sqrt{2}} (e^{i\eta}f_+ + e^{i\eta}f_-)$ and  $q =
\frac1{\sqrt{2}} (-ie^{i\eta}f_+ + i e^{i\eta}f_-)$. The Bogoliubov
Hamiltonian assumes a particularly simple form in these variables.
Defining velocities $v_j$ via $m v^2_j=[(2\beta-\gamma)n_0 - 2\beta
n_0 \cos(2\eta+\frac{2\pi j}{3})]$, or equivalently
\begin{equation}
v_j^2 = \frac{4\pi n_0\hbar^2}{5m^2}\left [-(a_0-a_4) +
\frac{10}{7}(a_2-a_4)\cos(2\eta +\frac{4\pi}{3}j)\right ].
\label{velocity}
\end{equation}
Then we have
\begin{eqnarray}\nonumber
{\cal H}_{B} &=& \sum_{{\bf
k}}\sum_{j=1}^{3}[mv^2_j+\frac{\hbar^2{\bf k}^2}{2m}]b^\dag_{j{\bf
k}}b_{j{\bf k}} -\frac{1}{2} mv^2_j(b^\dag_{j{\bf k}}b^\dag_{j{\bf
-k}}+b_{j{\bf k}}b_{j{\bf -k}})\\
&+&  [(\alpha+\gamma)n_0+\frac{\hbar^2{\bf k}^2}{2m}]p^\dag_{{\bf
k}}p_{{\bf k}} + \frac{(\alpha+\gamma)n_0}2(p^\dag_{{\bf
k}}p^\dag_{{\bf -k}}+p_{{\bf k}}p_{{\bf -k}})\nonumber\\
&+&  [-\gamma n_0+\frac{\hbar^2{\bf k}^2}{2m}]q^\dag_{{\bf
k}}q_{{\bf k}} + \frac{\gamma n_0}2(q^\dag_{{\bf k}}q^\dag_{{\bf
-k}}+q_{{\bf k}}q_{{\bf -k}}).
\end{eqnarray}
We see that these modes are completely decoupled and readily
diagonalized by the Bogoliubov transformation. Physically, the $b_j$
modes are connected to spin rotations about the three coordinate
axes with the spectrum $\omega_j(k) = k\sqrt{v^2_j+\hbar^2k^2/4m^2}$
($j=1,\,2,\,3$). The $p$ boson generates phase fluctuations of the
condensate and hence has the spectrum $\omega_4(k) =
k\sqrt{(\alpha+\gamma)n_0/m+\hbar^2k^2/4m^2}$ which involves the
compressibility $n_0(\alpha+\gamma)$. Finally the $q$ mode
corresponds to fluctuations of the $\eta$ parameter and has the
spectrum $\omega_5(k) = k\sqrt{(-\gamma)n_0/m+\hbar^2k^2/4m^2}$.
Note, in the long wavelength limit all these modes have a linear
dispersion. However, for a generic nematic superfluid, only the
first four are goldstone modes of the broken spin and phase
symmetry; the fifth  will actually acquire a gap once the degeneracy
between the different $\eta$ configurations is removed. The
contribution of these modes to the free energy is readily
calculated. However, only the dispersion of the $b_i$ modes depends
on the nematic parameter $\eta$ and will be responsible for lifting
the accidental degeneracy. The contribution from these modes $
\frac{\Delta F(\eta)}{V}=T\sum_{n=1}^{3}\int
{\frac{d^3\mathbf{k}}{(2\pi)^3}
\log(2\sinh\frac{\hbar\omega_n(k)}{2T}}) $ where $V$ is the volume,
will be discussed below.

{\it Order by Disorder:}
\begin{figure}
\includegraphics[width=.4\textwidth ]{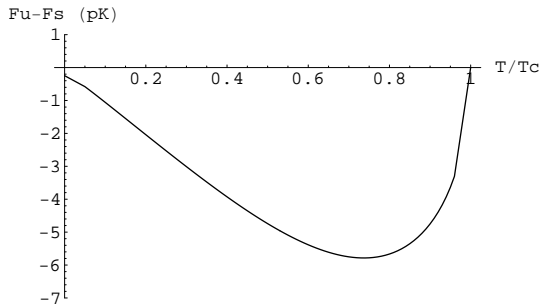}
\caption{Evolution of the free energy
difference between uniaxial and square biaxial nematic states as a
function of temperature with $^{87}$Rb parameters.}
\label{fig:Freeenergy}
\end{figure}
Before evaluating the free energy,  we make some general
observations arising from symmetry. For this purpose the nematic
region of the phase diagram is best described in terms of the
variables $(a_0-a_4)$, the $x$-axis in Fig.~\ref{fig:peace}, and the
ratio $\sigma = \frac{10}{7}\frac{a_4-a_2}{a_0-a_4}$. The nematic
region corresponds to $(a_0-a_4)<0$ and $|\sigma| <1$. Inspecting
the velocities Eq.~(\ref{velocity}) one sees that the degeneracy
remains unresolved along the line $\sigma=0$, which is also the
phase boundary for the full Hamiltonian due to the enlarged
symmetry. The symmetry breaking grows on moving away from this axis,
but has opposite effects on either side of this line. Formally, the
transformation $\sigma \rightarrow -\sigma,\,\eta \rightarrow
\eta+\pi/2$ leaves the approximate free energy invariant.

At zero temperature, the free energy reduces to the
contribution to the ground state energy for different nematic states
from the zero point motion of the harmonic modes. Remarkably, this
energy splitting $\Delta E(\eta)$ may be evaluated in closed form up
to an overall constant \footnote{In fact, this integral is divergent
at large $k$ because we have left off $\eta$-independent terms which
arise from the more methodical analysis \cite{us}}:

\begin{eqnarray}\nonumber
\frac{\Delta E(\eta)}{V}&=&\frac{8m^4}{15\pi^2\hbar^3}\sum_{j=1}^{3} v_j^5\\
&=&\frac{8\hbar^2}{15\pi^2m}\left [\frac{4\pi n_0(a_4-a_0)}{5}\right
]^{\frac{5}{2}} g_{\frac{5}{2}}(\eta,\,\sigma)
\label{eq:quantumdiff}
\end{eqnarray}
where it is convenient to define the series of functions:
$g_m(\eta,\, \sigma)= \sum_{j=1}^3 [1+\sigma\cos (2\eta+\frac{2\pi
j}{3})]^m$. The ground state is the one with the lowest combination
of velocities, and turns out to be the square biaxial nematic state
($\eta=\pi/2$) for $\sigma>0$, and the uniaxial state for $\sigma<0$
($\eta=0$). The latter is relevant to the experimental systems
spin-2 $^{87}$Rb and $^{23}$Na, which both are believed to have $\sigma
\approx -1$ \footnote{$\sigma=-1$ is an exact result if the scattering is
represented entirely by the spin exchange of the electrons.}.

\begin{figure}
\includegraphics[width=.4\textwidth]{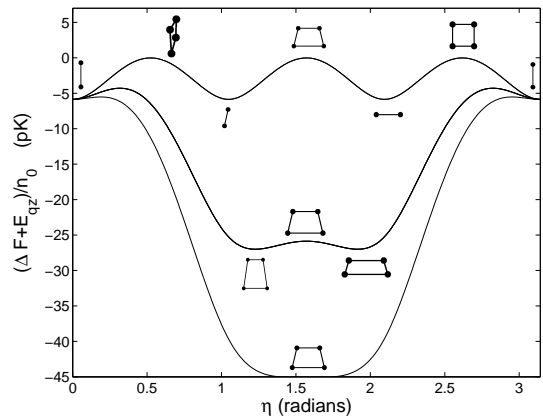}
\caption{Evolution of the uniaxial nematic into square
biaxial-nematic in a field. The free-energy, with a constant
subtracted, is plotted in $pK$ per atom for the conditions of Ref
\cite{schmaljohann04}, for $B=0,20,27$mG (top to bottom). Reciprocal
spinors are also shown. At the minima, they deform continuously from
rectangles into squares as the field is increased. The Ising-like
symmetry breaking at the critical field corresponds to making a
choice between which way to deform the square.}
\label{fig:quadraticZeeman}
\end{figure}

At finite temperatures, thermal fluctuations are much more effective
at lifting this degeneracy; again the state with the smaller
combination of velocities and hence a larger population of thermal
excitations will be entropically favored. While this may be
evaluated numerically, there is a
broad range of temperatures $T_c\gg T\gg {\rm max}(mv_i^2)$ where it
greatly simplifies. The lower limit is a rather small scale, related
to the magnetic energy and about $1$nK for $^{87}$Rb in
\cite{schmaljohann04}. In this limit, the linearly dispersing modes
obey the equipartition law, and the leading term in the free energy
that lifts the degeneracy is \footnote{Potentially larger terms,
proportional to $T^{\frac{5}{2}}$ and $T^{\frac{3}{2}}$ which
involves the functions $g_1(\eta,\sigma)$ do not have any $\eta$
dependence due to the sum over the three spin wave modes.}:
\begin{eqnarray} \nonumber
\frac{\Delta F(\eta)}{V}&=&-\frac{2}{3\pi}k_BT\sum_j(\frac{mv_j}{\hbar})^3\\
&=& -\frac{2}{3\pi}k_B T \left [\frac{4\pi n_0(a_4-a_0)}{5}\right
]^{\frac{3}{2}} g_{\frac32}(\eta,\sigma)
\end{eqnarray}
On evaluating this expression, thermal fluctuations are found to
lead to the same ground states as quantum fluctuations. A plot of
the free energy variation with $\eta$ for $\sigma \approx -1$ is in
the top curve of figure \ref{fig:quadraticZeeman}.

At higher temperatures the depletion of the condensate becomes
important. This may be readily taken into account by using the
temperature dependent condensate density obtained from mean field
theory, which for a harmonic trap is $n(T)=n_0[1-(T/T_c)^3]$. This
temperature dependence has a significant effect only near $T_c$ and
leads to a vanishing of the above free-energy splitting on
approaching $T_c$, assuming a continuous transition. This
temperature dependence of the free energy splitting between uniaxial
and square biaxial states is shown in Fig.~\ref{fig:Freeenergy}.
for the parameters of $^{87}$Rb in \cite{schmaljohann04}.

These results may be caricatured by an effective Landau theory,
where sixth order terms are generated by fluctuations. While a full
discussion of all such terms is undertaken elsewhere \cite{us}, we
note here that writing the five component order parameter as a
symmetric traceless matrix $\chi$ \cite{mermin74}, allows for a 6th
order term $\frac{32A}{3} \rm{tr}[\chi^3\chi^{*3}]$ in the free energy.
This term, when evaluated in the nematic subspace yields
$A|\chi_0|^6\cos 6\eta $, where $\chi_0$ is the magnitude of the
order parameter. Clearly, if $A>0$, then the free energy is
minimized by the square biaxial state $\eta = \pi/2, \,\pi/6,\,
5\pi/6$, while if $A<0$, then the free energy is minimized by the
uniaxial state $\eta = 0, \,\pi/3,\, 2\pi/3$. This sixth order term
vanishes rapidly on approaching the transition, accounting for the
near degeneracy of nematic ordered states near this point. Finally,
it may be recalled that biaxial nematics are rather difficult to
obtain in liquid crystal systems. Within Landau theory this is
explained by the presence of a cubic term in the free energy $B\tr
[\chi_R^3]$ when the nematic order parameter is a real symmetric
traceless matrix. This can be shown to favor the uniaxial state
regardless of the sign of $B$. However, in
the spinor condensate, the fact that the spin 2 field also carries a
charge quantum number implies the absence of such a term  (it is a
complex symmetric matrix with phase symmetry). Hence biaxial
nematics may be realized, and would naturally occur in the
$\sigma>0$ part of the nematic phase region. In addition to the
topological defects with non-abelian homotopy for the
general biaxial nematic, the square biaxial nematic realized here
also has half superfluid vortices bound to a spin defect.

\begin{figure}

\includegraphics[width=.47\textwidth]{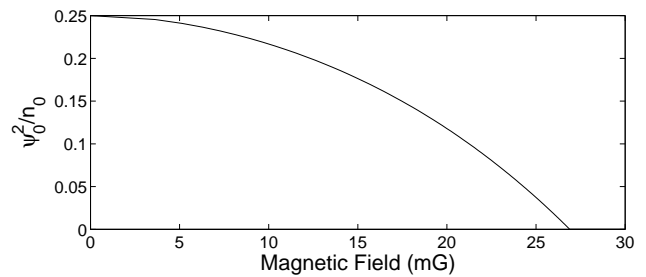}
\caption{Proportion of $F_z=0$ component
as magnetic field is increased to the critical field,
 for rubidium at density
$4\times10^{14}\mathrm{cc}^{-1}$ and $T=100\mathrm{nK}$.}
\label{fig:nematicization}
\end{figure}

{\it Experimental Prospects:} Magnetic fields make the phase diagram
more complex because the splitting due to fluctuations receives
competition from the quadratic Zeeman energy $H_{qz}=-cB^2\int
d^3\mathbf{x}{\psi^{\dagger}F_z^2\psi}$ \footnote{For a spinor
condensate, the linear Zeeman effect is unimportant since the total
magnetization is conserved on experimental time scales}, which
favors the square uniaxial state. For weak fields the nematic axis
drops into the $xy$ plane $\eta = \pi/3,\,2\pi/3$ because
$\ket{F_x=0}=(\sqrt{\frac{3}{8}},0,\frac{1}{2},0,\sqrt{\frac{3}{8}})$
has lower $H_{qz}$ than $\ket{F_z=0}$. On increasing the field, the
state continuously approaches the square state $\eta=\pi/2$.  In
\cite{schmaljohann04} a 50-50 split between atoms with $F_z=\pm 2$
was observed consistent with a square state; we believe this to be
due to the relatively large background field reported in the
experiments that overcomes the order by disorder effect. To observe
the latter effect, a lower magnetic field as estimated below is
needed. A possible way to observe the uniaxial to square state
transition is by measuring the fraction of atoms with $F_z=0$ along
a quantization axis parallel to the magnetic field as it drops from
25\% for uniaxial states to zero for the square state (See Fig.
\ref{fig:nematicization}). Magnitudes of fluctuation induced free
energy splittings  are shown in Table \ref{table:numbers} for
$^{87}$Rb and $^{23}$Na both at zero temperature and at $100$nK
(roughly half the critical temperature for the density in
\cite{schmaljohann04}). The critical field beyond which the
quadratic Zeeman effect dominates is shown in the last column. Given
that magnetic fields as low as $50\mathrm{mG}$ are readily
accessible in current experiments \cite{sadler06}, we believe these
effects are within experimental reach. The effect could also
possibly be used to access the condensate temperature by measuring
the critical field. In a trap, an important consideration is that
the linear part of the spin wave dispersions must be above the
quantization scale of the trap, $\hbar\omega_{\mathrm{trap}}<m \;
\mathrm{max}(v_i^2)$.



\begin{table}
\begin{tabular}{c|r|r|r|}
& $\Delta E$ &$\Delta F$ & $B_c$\\ \hline $^{87}$Rb & $0.3$ pK & $6$
pK &$27$ mG\\ \hline $^{23}$Na & $3$ pK & $9$ pK & $19$ mG\\ \hline
\end{tabular}
\caption{Free energy splittings at $T=0$, $T=100$nK, and critical
fields, calculated for $n=4\times 10^{14}\mathrm{cc}^{-1}$ - the
experimental conditions in Ref. \cite{schmaljohann04}. Sodium has a
greater disparity in its scattering lengths but has a stronger
quadratic Zeeman effect, explaining its lower critical field.}
\label{table:numbers}

\centering\end{table}
In conclusion, the role of fluctuations in selecting the ground
state spin structure of a spin-2 nematic condensate were studied.
Both quantum and thermal fluctuations were shown to lead to the
phase diagram in Fig.~\ref{fig:peace}.  For the experimentally
interesting cases of $^{23}$Na and $^{87}$Rb, these effects compete
against the quadratic Zeeman term, but can predominate for
sufficiently weak fields. An Ising transition marks this point,
which we believe can be accessed readily in future experiments. If
so, this would be one of the first experimental demonstrations of
the `order-by-disorder' effect widely studied in the context of
frustrated quantum magnetism.

We would like to thank Daniel Podolsky, Gil Refael and Dan
Stamper-Kurn for insightful discussions and Fei Zhou for alerting us
to \cite{zhou_fluct}, where similar results are obtained. Support
from from the Hellman Family Fund and LBNL DOE-504108 (A.V.),  NSF
Career Award, AFOSR and Harvard-MIT CUA (E.D. and A.M.T.) and the
Sherman Fairchild Foundation (R.B.) are acknowledged.

\bibliography{spin3}
\end{document}